\documentclass[traditabstract]{aa}

\usepackage{natbib} 
\bibpunct{(}{)}{;}{a}{}{,} 
\usepackage{graphicx}
\usepackage{amsmath}
\usepackage{txfonts}
\usepackage{xspace}
\usepackage{amssymb}
\newcommand{\rhk}{$\log R'_{\rm HK}$\xspace}

\newcommand{\feh}{\mbox{[Fe/H]}\xspace}
\newcommand{\teff}{\mbox{$T_{\rm *, eff}$}\xspace}
\newcommand{\logg}{\mbox{$\log g$}\xspace}
\newcommand{\vsini}{\mbox{$v \sin i_{*}$}\xspace}
\newcommand{\mictrb}{\mbox{$\xi_{\rm t}$}\xspace}
\newcommand{\mactrb}{\mbox{$v_{\rm mac}$}\xspace}

\newcommand{\kms}{\mbox{km\,s$^{-1}$}\xspace}

\newcommand{\halpha}{\mbox{$H_\alpha$}\xspace}

\newcommand{\mjup}{\mbox{$M_{\rm Jup}$}\xspace}
\newcommand{\rjup}{\mbox{$R_{\rm Jup}$}\xspace}
\newcommand{\densjup}{\mbox{$\rho_{\rm Jup}$}\xspace}

\newcommand{\rstar}{\mbox{$R_{*}$}\xspace}

\newcommand{\msol}{\mbox{$M_\odot$}\xspace}
\newcommand{\rsol}{\mbox{$R_\odot$}\xspace}

\newcommand{\svcos}{\mbox{$\sqrt{\vsini} \cos \lambda$}\xspace} 
\newcommand{\svsin}{\mbox{$\sqrt{\vsini} \sin \lambda$}\xspace}

\newcommand{\prot}{\mbox{$P_{\mathrm{rot}}$}\xspace}

\begin{document}
   \title{WASP-71b: a bloated hot Jupiter in a 2.9-day, prograde orbit around an evolved F8 star}

\authorrunning{A. M. S. Smith \textit{et al.}}
\titlerunning{WASP-71b}

\author{
	A.~M.~S.~Smith\inst{\ref{inst1},\ref{inst2}}\and
          D.~R.~Anderson\inst{\ref{inst1}}\and
          F.~Bouchy\inst{\ref{inst3},\ref{inst4}}\and
          A.~Collier Cameron\inst{\ref{inst5}}\and
          A.~P.~Doyle\inst{\ref{inst1}}\and
          A.~Fumel\inst{\ref{inst6}}\and
          M.~Gillon\inst{\ref{inst6}}\and
          G.~H\'{e}brard\inst{\ref{inst3},\ref{inst4}}\and
          C.~Hellier\inst{\ref{inst1}}\and
          E.~Jehin\inst{\ref{inst6}}\and
          M.~Lendl\inst{\ref{inst7}}\and
          P.~F.~L.~Maxted\inst{\ref{inst1}}\and
          C.~Moutou\inst{\ref{inst8}}\and
          F.~Pepe\inst{\ref{inst7}}\and
          D.~Pollacco\inst{\ref{inst9}}\and
          D.~Queloz\inst{\ref{inst7}}\and
          A.~Santerne\inst{\ref{inst8},\ref{inst4}}\and
          D.~Segransan\inst{\ref{inst7}}\and
          B.~Smalley\inst{\ref{inst1}}\and
          J.~Southworth\inst{\ref{inst1}}\and
          A.~H.~M.~J.~Triaud\inst{\ref{inst7}}\and
          S.~Udry\inst{\ref{inst7}}\and
          R.~G.~West\inst{\ref{inst10}}
                    }

\institute{
	Astrophysics Group, Lennard-Jones Laboratories, Keele University, Keele, Staffordshire, ST5 5BG, UK\label{inst1} \and
	N. Copernicus Astronomical Centre, Polish Academy of Sciences, Bartycka 18, 00-716 Warsaw, Poland. email:amss@camk.edu.pl\label{inst2}\and
	Institut d'Astrophysique de Paris, UMR7095 CNRS, Universit\'{e} Pierre \& Marie Curie, 98bis boulevard Arago, 75014, Paris, France\label{inst3}\and
	Observatoire de Haute-Provence, CNRS/OAMP, 04870 Saint-Michel-l'Observatoire, France\label{inst4}\and
	SUPA, School of Physics and Astronomy, University of St. Andrews, North Haugh, St. Andrews, Fife, KY16 9SS, UK\label{inst5}\and
	Institut d'Astrophysique et de G\'{e}ophysique, Universit\'{e} de Li\`{e}ge, All\'{e}e du 6 Ao\^{u}t 17, Sart Tilman, Li\`{e}ge 1, Belgium\label{inst6}\and
	Observatoire de Gen\`{e}ve, Universit\'{e} de Gen\`{e}ve, 51 Chemin des Maillettes, 1290 Sauverny, Switzerland\label{inst7}\and
	 Aix Marseille Universit\'e, CNRS, LAM (Laboratoire d'Astrophysique de Marseille) UMR 7326, 13388, Marseille, France\label{inst8}\and
	Department of Physics, University of Warwick, Coventry CV4 7AL, UK\label{inst9}\and
	Department of Physics \& Astronomy, University of Leicester, Leicester, LE1 7RH, UK\label{inst10}
             }

   \date{Received November 13, 2012; accepted March 14, 2013}

\abstract{We report the discovery by the WASP transit survey of a highly-irradiated, massive ($2.242 \pm 0.080$~\mjup) planet which transits a bright ($V=10.6$), evolved F8 star every 2.9 days. The planet, WASP-71b, is larger than Jupiter ($1.46\pm0.13$~\rjup), but less dense ($0.71\pm0.16$~\densjup). We also report spectroscopic observations made during transit with the CORALIE spectrograph, which allow us to make a highly-significant detection of the Rossiter-McLaughlin effect. We determine the sky-projected angle between the stellar-spin and planetary-orbit axes to be $\lambda = 20.1\pm9.7$~degrees, i.e. the system is `aligned', according to the widely-used alignment criteria that systems are regarded as misaligned only when $\lambda$ is measured to be greater than 10~degrees with 3-$\sigma$ confidence. WASP-71, with an effective temperature of $6059 \pm 98$~K, therefore fits the previously observed pattern that only stars hotter than 6250~K are host to planets in misaligned orbits. We emphasise, however, that $\lambda$ is merely the sky-projected obliquity angle; we are unable to determine whether the stellar-spin and planetary-orbit axes are misaligned along the line-of-sight. With a mass of $1.56\pm0.07$~\msol, WASP-71 was previously hotter than 6250~K, and therefore might have been significantly misaligned in the past. If so, the planetary orbit has been realigned, presumably through tidal interactions with the cooling star's growing convective zone.
}


   \maketitle
%

\section{Introduction}

The recent  {\it Kepler} discoveries of thousands of transiting exoplanet candidates (e.g. \citealt{Batalha13}) has extended the parameter space of planet discovery and made major advances in answering questions about the statistical population of planetary systems in our Galaxy (e.g. \citealt{Howard12}).

The vast majority of the systems detected by {\it Kepler}, however, are too faint for in-depth characterisation, which is required to increase our knowledge of the range of properties exhibited by the nearby planetary population, as well as to make advances in our understanding of planetary formation and evolution. Examples of planet characterisation that require bright targets include radial velocities to measure the star-to-planet mass ratio, orbital eccentricity and spin-orbit alignment and transit and occultation spectra and photometry at a range of wavelengths to measure the planet's atmospheric properties.

Discovering relatively bright transiting systems that are amenable for characterisation measurements is where ground-based transit surveys such as WASP \citep{Pollacco06} and HAT \citep{bakos02} are vital. Between them, these two surveys have discovered more than 100 systems around bright stars ($8.5 \lesssim V \lesssim 12.5$); a sample which comprises the majority of the systems suitable for in-depth characterisation.

One of the characterisation measurements that can be made is spectroscopic transit measurements with the aim of detecting the changes in apparent radial velocity characteristic of the Rossiter-McLaughlin (RM) effect \citep{Holt1893,Rossiter,McLaughlin}. As the planet obscures a portion of the star rotating towards us, an anomalous red-shift is observed in the radial velocities of the star, and a blue shift is observed during the planet's passage across the portion of the star rotating away from us. The shape of the RM signature is therefore sensitive to the path of the planet across the stellar disk relative to the stellar spin axis. It is possible to determine the sky-projection of the obliquity angle, $\lambda$, between the stellar-rotation axis and the planetary-orbital axis.

Observations of the RM effect have been made for around fifty planetary systems\footnote{Ren\'{e} Heller's Holt-Rossiter-McLaughlin Encyclopaedia at http://www.aip.de/People/RHeller}, revealing a significant fraction of them to be misaligned. This discovery was unexpected, since the proto-planetary disc from which planets form is expected to rotate in the same direction as the star, and the inward migration of hot Jupiters to their current locations via tidal interaction with the gas disc is expected to preserve this alignment \citep{Lin96}.

As a result of these observations, planet-planet interactions are favoured as the dominant mechanism for producing hot Jupiters \citep{Morton&Johnson11}, resulting in a large number of misaligned systems. Those planets which orbit relatively cool dwarfs (\teff $<$ 6250~K) are then probably quickly re-aligned through tidal interactions with the large stellar convective zone, resulting in an apparent correlation between misalignment and stellar effective temperature \citep{Winn10}.

Here we report the discovery of a transiting planet orbiting the $V = 10.6$ star WASP-71 (=TYC 30-116-1) in the constellation Cetus, and a detection of the Rossiter-McLaughlin effect in the system.

\section{Observations}
\subsection{WASP photometry}

WASP-71 was observed in 2008 and 2009 by WASP-South, which is located at the South African Astronomical Observatory (SAAO), near Sutherland in South Africa, and in 2008, 2009 and 2010 by SuperWASP-N at the Observatorio del Roque de los Muchachos on La Palma, Spain. Each instrument consists of eight Canon 200mm f/1.8 lenses, each equipped with an Andor $2048\times2048$ e2v CCD camera, on a single robotic mount. Further details of the instrument, survey and data reduction procedures are described in \cite{Pollacco06} and details of the candidate selection procedure can be found in \cite{Cameron-etal07} and \cite{wasp3}. A total of 9316 measurements of WASP-71 were made using WASP-South from 2008 July 30 to 2009 December 12, and 11166 using SuperWASP-N from 2008 September 24 to 2010 December 27, making a grand total of 20482 observations.

The data revealed the presence of a transit-like signal with a period of $\sim2.9$ days and a depth of $\sim4$ mmag. The combined WASP light curve is shown folded on the best-fitting orbital period in the upper panel of Figure~\ref{fig:phot}.

\subsection{Spectroscopic follow-up}

Spectroscopic observations were conducted using the SOPHIE spectrograph mounted on the 1.93-m telescope of the Observatoire de Haute-Provence, France. Nine observations were made between 2011 July 29 and 2011 August 26. Further spectroscopic observations were carried out using the CORALIE spectrograph of the 1.2-m Euler-Swiss telescope at La Silla Observatory, Chile. A total of 31 observations were acquired using CORALIE between 2011 September 05 and 2011 November 28. Sixteen of these were made on 2011 September 27 (12 during transit), with the intention of detecting the Rossiter-McLaughlin effect. 

SOPHIE was used in high-resolution ($R = 75000$) mode, with the CCD in fast-readout mode. A second fibre was placed on the sky to check for sky background contamination in the spectra of the target. More information about the SOPHIE instrument can be found in \cite{sophie}. The CORALIE observations were conducted only in dark time to avoid moon light entering the fibre, and the instrumental FWHM was $0.11 \pm 0.01\AA$.

Observations of thorium-argon emission line lamps were used to calibrate both the SOPHIE and CORALIE spectra. The data were processed using the standard SOPHIE and CORALIE data-reduction pipelines. The resulting radial velocity data are listed in Table \ref{tab:rv} and plotted in Figure \ref{fig:rv}. In order to rule out the system as a blended eclipsing binary, we examined the bisector spans (BS) (e.g. \citealt{Queloz01}), which exhibit no significant correlation with radial velocity (Figure \ref{fig:rv}, lower panel), as expected for a true planetary system.

\begin{table}[h]
\centering
\caption{Radial velocities of WASP-71}
\begin{tabular}{cccrc} \hline
Time & RV & $\sigma_{\mathrm RV}$ & \multicolumn{1}{c}{BS} & S/N$^\dagger$\\
BJD(UTC)& km s$^{-1}$ &  km s$^{-1}$ & \multicolumn{1}{c}{km s$^{-1}$} & ...\\
 -- 2450000&&&&\\
 \hline
\multicolumn{5}{l}{SOPHIE:}\\

5771.6206 & 7.600 & 0.017 & -0.021 & 24.2\\
5772.6279 & 8.016 & 0.017 & -0.085 & 24.8\\
5773.6183 & 7.894 & 0.018 & -0.032 & 22.5\\
5774.6240 & 7.599 & 0.019 & 0.027 & 26.2\\
5793.6378 & 8.009 & 0.018 & 0.079 & 22.6\\
5794.6299 & 7.591 & 0.018 & 0.045 & 22.7\\
5795.6269 & 7.926 & 0.017 & -0.057 & 22.9\\
5797.6420 & 7.603 & 0.017 & 0.005 & 22.7\\
5799.6215 & 7.957 & 0.018 & -0.012 & 22.0\\
&&&&\\
\multicolumn{5}{l}{CORALIE:}\\
5809.7893 & 7.708 & 0.010 & -0.005 & 39.7\\
5810.8425 & 8.005 & 0.012 &  0.030 & 33.6\\
5811.8113 & 7.624 & 0.014 & -0.070 & 29.9\\
5812.9114 & 7.814 & 0.014 & -0.021 & 31.6\\
5813.8670 & 8.009 & 0.011 &  0.016 & 37.8\\
5814.8860 & 7.597 & 0.015 & -0.005 & 31.7\\
5825.7565 & 7.892 & 0.010 & -0.024 & 41.5\\
5826.8005 & 7.566 & 0.013 & -0.066 & 32.7\\
5828.8688 & 7.801 & 0.012 & -0.032 & 36.4\\
5829.8431 & 7.601 & 0.011 &  0.038 & 39.2\\
5830.8440 & 8.013 & 0.013 & -0.034 & 33.8\\
5831.6393 & 7.856 & 0.014 & -0.023 & 30.9\\
5831.6573 & 7.839 & 0.014 & -0.039 & 30.7\\
5831.6758 & 7.869 & 0.013 & -0.025 & 32.1\\
5831.6921 & 7.851 & 0.013 & -0.063 & 32.7\\
5831.7083 & 7.847 & 0.013 & -0.033 & 33.5\\
5831.7245 & 7.850 & 0.013 & -0.035 & 32.5\\
5831.7408 & 7.810 & 0.013 & -0.033 & 32.6\\
5831.7570 & 7.792 & 0.013 & -0.032 & 32.3\\
5831.7776 & 7.774 & 0.013 & -0.002 & 32.3\\
5831.7938 & 7.749 & 0.013 & -0.019 & 33.3\\
5831.8101 & 7.727 & 0.013 & -0.008 & 32.8\\
5831.8263 & 7.692 & 0.013 &  0.036 & 32.8\\
5831.8425 & 7.712 & 0.013 &  0.001 & 32.3\\
5831.8587 & 7.691 & 0.013 & -0.023 & 32.4\\
5831.8805 & 7.743 & 0.014 & -0.004 & 34.3\\
5831.8967 & 7.724 & 0.014 & -0.063 & 32.8\\
5833.7874 & 7.999 & 0.012 & -0.061 & 34.2\\
5858.7334 & 7.560 & 0.016 & -0.075 & 26.9\\
\hline
\\
\end{tabular}\\
$\dagger$ Signal-to-noise ratio per pixel at a wavelength of 550~nm.
\label{tab:rv}
\end{table}

\subsection{Photometric follow-up}

High-precision photometric follow-up observations were performed using the robotic Transiting Planets and Planetesimals Small Telescope (TRAPPIST; \citealt{TRAPPIST}) at La Silla. A complete transit was observed on 2011 September 27 (simultaneously with our spectroscopic transit observations), and a partial transit was observed on 2011 September 30, both in the $z$-band. The data were reduced using the {\sc iraf/daophot} package, with aperture radii and choice of comparison stars chosen to minimise the out-of-transit light curve rms. Both TRAPPIST light curves are shown in the lower panel of Figure \ref{fig:phot}.

\section{Determination of system parameters}
\subsection{Stellar parameters}
\label{sec:spec}

The 31 individual CORALIE spectra of WASP-71 were co-added to produce a single spectrum with an average S/N of around 100:1. The analysis was performed using the methods described in \citet{2009A&A...496..259G} and standard pipeline reduction products. We used the \halpha\ line to determine the effective temperature (\teff), and the surface gravity (\logg) was determined from the Ca~{\sc i} lines at 6162{\AA} and 6439{\AA} \citep{2010A&A...519A..51B}, along with the Na~{\sc i} D lines.

The elemental abundances were determined from equivalent width measurements of several clean and unblended lines. A value for microturbulence (\mictrb) was determined from Fe~{\sc i} using the method of \cite{1984A&A...134..189M}. The quoted error estimates include the contributions from the uncertainties in \teff, \logg\ and \mictrb, as well as the scatter due to measurement and atomic data uncertainties. 

The projected stellar rotation velocity (\vsini) was determined by fitting the profiles of several unblended Fe~{\sc i} lines. We assumed a value for macroturbulence (\mactrb) of 3.3 $\pm$ 0.3 {\kms}, based on the
calibration by \cite{2010MNRAS.405.1907B}, and an instrumental FWHM of 0.11 $\pm$ 0.01~{\AA} was determined from the telluric lines around 6300\AA. The results of the spectral analysis are listed in Table~\ref{tab:stellar}.

\subsubsection{Stellar activity}

We searched the WASP photometry for rotational modulations, using the sine-wave fitting algorithm described by \cite{w41}. We estimated the significance of periodicities by subtracting the fitted transit light curve and then repeatedly and randomly permuting the nights of observation. No significant periodic modulation was detected, to a 95 per cent confidence upper limit of 1 mmag.

The CORALIE spectra do not allow a measurement of \rhk because of the low signal-to-noise ratio of the blue end of the spectrum, and the presence of thorium arc lines. The SOPHIE spectra are also too noisy to make a meaningful measurement of \rhk, and there is no obvious emission observed in the cores of the Ca II lines.\\

\subsubsection{Stellar age}
\label{sec:age}

\begin{figure} 
\includegraphics[width=0.5\textwidth]{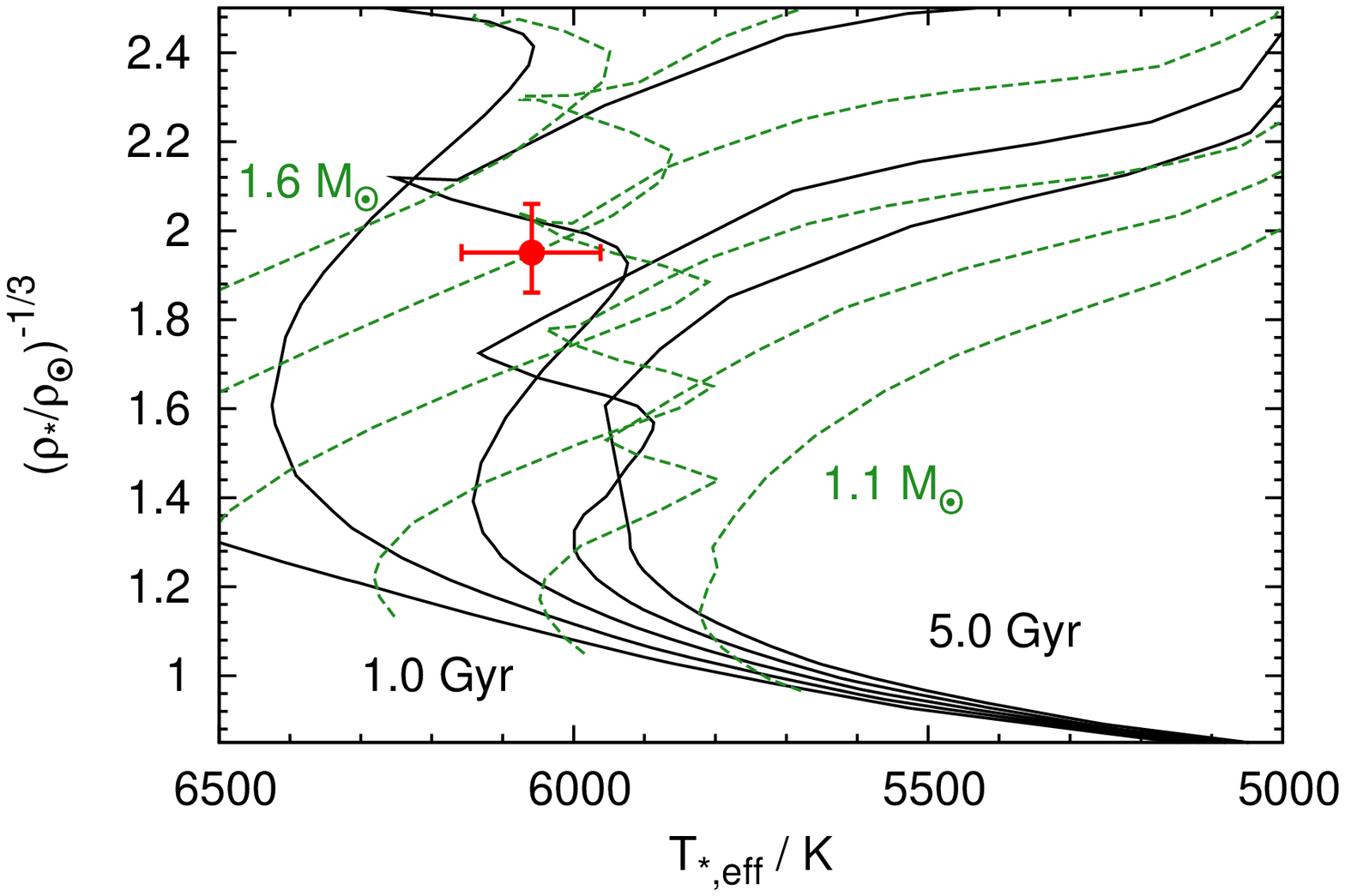} 
\caption{Modified Hertzsprung-Russell diagram. WASP-71 is plotted alongside isochrones and evolutionary tracks from the models of \cite{Marigo} for \feh = 0.14 ($Z = 0.026$ using the approximation $[Fe/H] = \log(Z/Z_\odot$, with $Z_\odot = 0.019$). The red circle corresponds to our best-fitting stellar density and stellar effective temperature. The isochrones (solid black lines)correspond (from left to right) to ages 1.0, 2.0, 3.0, 4.0, and 5.0 Gyr. The evolutionary tracks (dashed green lines) correspond (from lower to upper) to masses of 1.1, 1.2, 1.3, 1.4, 1.5 and 1.6 solar masses.}
\label{fig:hr} 
\end{figure}

The degree of lithium absorption in the spectrum (Table \ref{tab:stellar}) suggests that WASP-71 has an age of
$\sim$2~Gyr \citep{2005A&A...442..615S}. The measured stellar \vsini gives an upper limit to the rotation period, \prot, of $11.8 \pm 2.7$~d, using the best-fitting stellar radius (Table~\ref{tab:mcmc}). This corresponds to an upper limit on the age of $1.6^{+1.4}_{-0.8}$~Gyr using the gyrochronological relation of \cite{2007ApJ...669.1167B}. In Figure~\ref{fig:hr} we plot WASP-71 alongside the stellar evolution tracks of \cite{Marigo}, from which we infer an age of $3\pm1$~Gyr.  In conclusion, the precise stellar age is somewhat uncertain, but the three methods used give good agreement, and it seems likely the star is around 2 -- 3~Gyr old.

\subsubsection{Stellar distance}

We calculated the distance ($345 \pm 32$~pc) to WASP-71 using the apparent V-band magnitude of $10.56 \pm 0.09$ from the Tycho catalogue, and a stellar luminosity calculated from our best-fitting stellar effective temperature and radius (Table \ref{tab:mcmc}). We adopted a bolometric correction of $-0.1$ and do not correct for interstellar reddening, which we expect to negligible because of the lack of interstellar Na D lines in the CORALIE spectra.

\begin{table}[h]
\caption{Stellar Parameters and Abundances from Analysis of CORALIE Spectra}
\begin{tabular}{ll|cc} \hline
Parameter  & Value & 							Parameter & Value \\ \hline
RA (J2000.0) & 01h57m03.20s & 					{[Fe/H]}   & $+$0.14 $\pm$ 0.08 \\
Dec (J2000.0) & $+00^{\circ}~45\arcmin~31.9\arcsec $&	{[Na/H]}   & $+$0.34 $\pm$ 0.05 \\
\teff      &  6050 $\pm$ 100 K & 						{[Si/H]}   & +0.26 $\pm$ 0.12 \\
\logg (cgs)     & 3.9 $\pm$  0.15 & 					{[Ca/H]}   & +0.25 $\pm$ 0.16 \\
\mictrb\footnote{Microturbulent velocity}    &  1.5  $\pm$ 0.1 \kms & 
											{[Sc/H]}   & +0.14 $\pm$ 0.07 \\
\vsini     &  9.4 $\pm$ 0.5 \kms &		 			{[Ti/H]}   & +0.12 $\pm$ 0.18 \\
$\log A$(Li)  &    2.11 $\pm$ 0.09 & 					{[V/H]}    & +0.15 $\pm$ 0.09 \\
Sp. Type   &   F8 & 								{[Cr/H]}   & +0.11 $\pm$ 0.11 \\
Distance   &    345 $\pm$ 32 pc & 					{[Mn/H]}   & +0.18 $\pm$ 0.11 \\
Mass       &    1.42 $\pm$ 0.14 $\msol$&				{[Co/H]}   & +0.19 $\pm$ 0.07 \\  
Radius     &   2.20 $\pm$ 0.48 $\rsol$ &				{[Ni/H]}   & +0.19 $\pm$ 0.07 \\
\hline
\multicolumn{4}{l}{Additional identifiers for WASP-71:}\\
\multicolumn{4}{l}{TYC 30-116-1}\\
\multicolumn{4}{l}{2MASS J01570320+0045318}\\
\multicolumn{4}{l}{1SWASP J015703.20+004531.9}\\
\hline
\\
\end{tabular}
\label{tab:stellar}
\newline {\bf Note:} The spectral type was estimated from \teff\
using the table of \cite[p. 507]{gray-book}. The mass and radius were estimated using the \cite{2010A&ARv..18...67T} calibration.

\end{table}

\begin{figure} 
\includegraphics[width=11cm,angle=270]{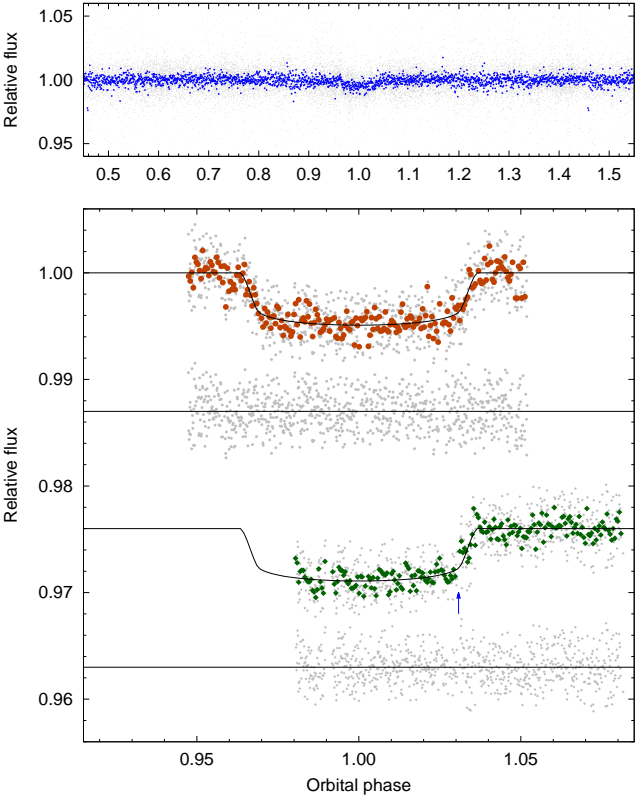} 
\caption{Photometry. {\it Upper panel:} Combined SuperWASP-N and WASP-South discovery light curve for WASP-71b, folded on the orbital period of $P = 2.9036747~d$. The small, light-grey points are the un-binned data and the larger blue points are binned in phase, with a bin width equivalent to 120~s.
{\it Lower panel:} High-precision transit photometry from TRAPPIST, over-plotted with our best-fitting models (solid lines). In each case the un-binned data are shown in small, light-grey points, with residuals to the best-fitting model shown below. Data binned on 120~s are indicated with orange circles for the transit of 2011 September 27 (upper) and green squares for the transit of 2011 September 30 (lower), which is offset in flux for clarity. The phase at which the meridian flip occurred during the latter transit is indicated by a blue arrow.
}
\label{fig:phot} 
\end{figure} 

\subsection{Planetary system parameters}

The WASP and TRAPPIST photometry were combined with the SOPHIE and CORALIE radial velocities and analysed simultaneously using the Markov Chain Monte Carlo (MCMC) method. One of the SOPHIE radial velocity measurements was made during transit, and we exclude this data point from our analysis (but show it as a grey point in Figure \ref{fig:rv} for completeness), lest it bias the spectroscopic transit data from CORALIE. We use the current version of the MCMC code described in \cite{Cameron-etal07} and \cite{wasp3}, which uses the relation between stellar mass, effective temperature, density and metallicity appropriate for stars with masses between 0.2 and 3.0 \msol from \cite{SWorth_homo4}. Briefly, the radial velocity data are modelled with a Keplerian orbit and the RM effect is modelled using the formulation of \cite{Gimenez_RM}. The photometric transits are fitted using the model of \cite{M&A} and limb-darkening was accounted for using a four-coefficient, non-linear model, employing coefficients appropriate to the passbands from the tabulations of \cite{Claret, Claret04}. The coefficients were determined using an initial interpolation in $\log g_{*}$ and [Fe/H] (values from Table \ref{tab:stellar}), and an interpolation in $T_{\rm *, eff}$  at each MCMC step. The coefficient values corresponding to the best-fitting value of $T_{\rm *, eff}$ are given in Table \ref{tab:limb}.

The MCMC proposal parameters we use are: the epoch of mid-transit, $T_{\rm c}$; the orbital period, $P$; the transit duration, $T_{\rm 14}$; the fractional flux deficit that would be observed during transit in the absence of stellar limb-darkening, $\Delta F$; the transit impact parameter for a circular orbit, $b = a\cos(i_P)/\rstar$; the stellar reflex velocity semi-amplitude, $K_{\mathrm{*}}$; the stellar effective temperature, \teff, and the stellar metallicity, [Fe/H]. When fitting for a non-zero orbital eccentricity, there are an additional two proposal parameters: $\sqrt{e}\cos\omega$ and $\sqrt{e}\sin\omega$, where $e$ is the orbital eccentricity, and $\omega$ is the argument of periastron \citep{wasp30}. A further two proposal parameters are included when modelling the RM effect, these are \svcos and \svsin, where \vsini is the sky-projected stellar rotation velocity, and $\lambda$ is the sky-projected obliquity (the angle between the axes of stellar rotation and planetary orbit).

The best-fitting system parameters are taken to be the median values of the posterior probability distribution. Linear functions of time were fitted to each light curve at each step of the MCMC, to remove systematic trends.   Given that we have a measurement of \vsini from our spectral analysis, we place a Gaussian prior on the value of \vsini within the MCMC, by means of a Bayesian penalty on $\chi^2$.

An initial MCMC fit for an eccentric orbit found $e = 0.0253\pm0.0095$ ($\omega = 136^{+29}_{-296}$ degrees), with a 3-$\sigma$ upper-limit to the eccentricity of 0.060, but this is consistent with a circular orbit. Following the F-test approach of \cite{lucy_sweeney}, we find that there is a 9.5 per cent probability that the apparent eccentricity could have arisen if the underlying orbit were actually circular. Given this, and also that there are good reasons, theoretical and empirical, to expect the orbits of short-period planets such as WASP-71b to be circular (e.g. \citealt{w44_45_46}), we fix $e=0$ in subsequent analyses. We note that the values of the other model parameters, and their associated uncertainties, are almost identical to those of the eccentric orbit solution.

We also tried fitting for a linear trend in the RVs with the inclusion of an additional parameter in our MCMC fit. Such a trend (such as that found in the RVs of WASP-34, \citealt{wasp34}) may indicate the presence of a third body in the system. The best-fitting linear radial acceleration ($d \gamma / dt = 12.7^{+5.8}_{-4.2}$~m~s$^{-1}$~yr$^{-1}$) is consistent with zero, indicating there is no evidence for an additional body in the system based on our RVs, which span 122 d. The orbital parameters we report are the result of a fit which does not allow for a linear trend in radial velocity.

The system parameters derived from our best-fitting circular-orbit model are presented in Table \ref{tab:mcmc}. The corresponding transit and RV models are superimposed on our data in Figures \ref{fig:phot} and \ref{fig:rv} respectively.

\subsection{Spectroscopic transit}

\begin{figure} 
\includegraphics[width=0.5\textwidth]{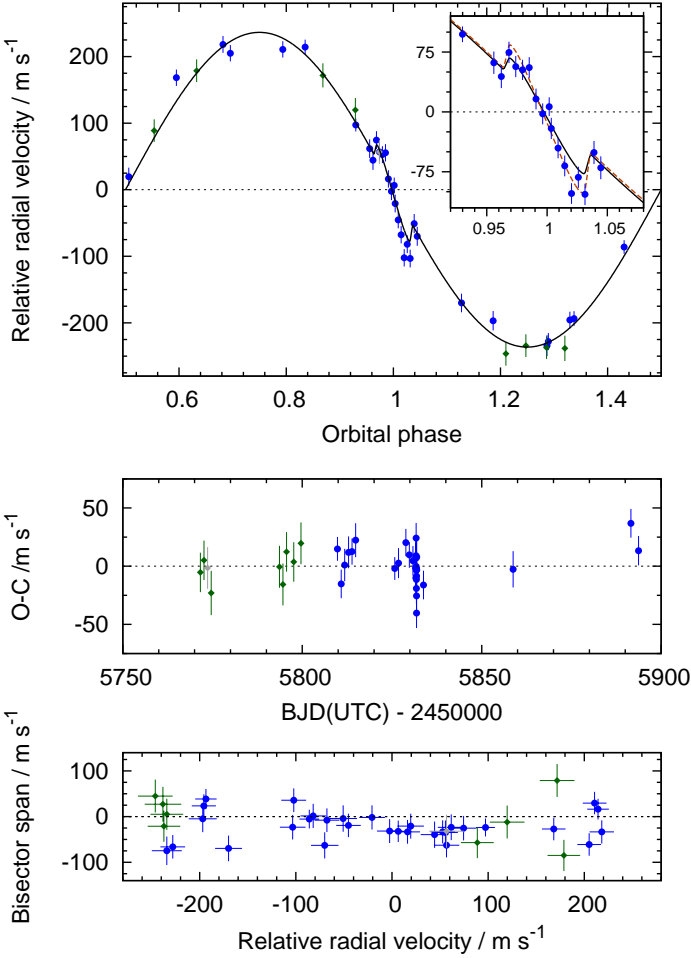} 
\caption{Radial velocities. {\it Upper panel:} Phase-folded radial-velocity measurements from SOPHIE (green squares) and CORALIE (blue circles). The SOPHIE measurement that fell during transit and is excluded from our MCMC analysis is shown as a grey square. The centre-of-mass velocity, $\gamma$ = -7.7941 km s$^{-1}$, has been subtracted, as has the fitted offset $\gamma_{\rm COR-SOPH} = 41.6$~m~s$^{-1}$ between the two datasets. The best-fitting MCMC solution is over-plotted as a solid black line. {\it Inset:} A close up of the Rossiter-McLaughlin effect. In addition to the model plotted in the main panel, a second model where $\vsini$ is not constrained with a prior is shown with a dashed orange line.
{\it Middle panel:} Residuals from the Keplerian and RM fit as a function of time. No significant linear trend is present in the residuals (the best-fitting linear radial acceleration is $d \gamma / dt = 12.7^{+5.8}_{-4.2}$~m~s$^{-1}$~yr$^{-1}$).
{\it Lower panel:} Bisector span measurements as a function of radial velocity. The uncertainties in the bisectors are taken to be twice the uncertainty in the radial velocities.
}
\label{fig:rv} 
\end{figure}

The Rossiter-McLaughlin effect is clearly detected in the radial velocities taken with CORALIE during transit (Fig.~\ref{fig:rv}); we find $\lambda = 20.1 \pm9.7$~deg. The best-fitting $\vsini = 9.89 \pm 0.48$~km~s$^{-1}$ is constrained by the Bayesian prior ($9.4\pm 0.5$~km~s$^{-1}$) that we imposed. If left unconstrained, the data favour a significantly higher value of $\vsini = 17.6\pm2.1$~km~s$^{-1}$, which is incompatible with our spectroscopic analysis (Sec. \ref{sec:spec}). There is often a degeneracy between \vsini and \mactrb (e.g. \citealt{w36}), but even if \mactrb were zero, the spectroscopic \vsini would only be $\approx0.5$~km~s$^{-1}$ higher.

The model from our unconstrained MCMC analysis, which is over-plotted with our standard model in the inset of the uppermost panel of Fig.~\ref{fig:rv},  is a better fit to the radial velocities, particularly those near the end of transit. The values of the other system parameters reported in Table~\ref{tab:mcmc} are unchanged, as they also are when the spectroscopic transit data are omitted from the analysis and the only model fitted to the remaining RVs is a Keplerian orbit.

\begin{table*} 
\caption{System parameters} 
\label{tab:mcmc}
\begin{tabular}{lccc}
\hline
\hline
Parameter & Symbol & Unit & Value\\
\hline 
\textit{Model parameters:} &\\
&\\
Orbital period	    	    	    	    & 	$P$ & d & $2.9036747\pm0.0000068$\\
Epoch of mid-transit	    	    	    & 	$T_{\rm c}$ &HJD, UTC & $2455738.84974\pm0.00070$\\
Transit duration    	    	    	    & 	$T_{\rm 14}$ &d & $0.2091\pm0.0019$\\
Planet-to-star area ratio   	    	    & 	$\Delta F=R_{\rm P}^{2}$/R$_{*}^{2}$&... & $0.00443\pm0.00015$\\
Transit impact parameter    	    & 	$b$ &...& $0.39\pm0.14$\\
Stellar orbital velocity semi-amplitude     & 	$K_{\rm *}$ &km s$^{-1}$ & $0.2361\pm0.0037$\\
System velocity     	    	    	    &     	$\gamma$ &km s$^{-1}$ & $7.7941\pm0.0015$\\
Velocity offset between CORALIE and SOPHIE & $\gamma_{\rm COR-SOPH}$ & m s$^{-1}$ & $41.6 \pm 1.6$\\
Stellar effective temperature         &  $T_{\rm *, eff}$ & K & $6059\pm98$\\
Stellar metallicity			    &  [Fe/H] & dex & $+0.140\pm0.080$\\
...						    & \svcos & km$^{1/2}$ s$^{-1/2}$& $2.95^{+0.16}_{-0.21}$\\
...						    & \svsin  & km$^{1/2}$ s$^{-1/2}$& $1.08\pm0.50$\\
&\\
\textit{Derived parameters:}\\
&\\
Ingress / egress duration    	    & 	$T_{\rm 12}=T_{\rm 34}$ &d & $0.0152\pm0.0023$\\
Orbital inclination angle   	    	    & 	$i_P$ &$^\circ$  & $84.9\pm2.2$\\
Orbital eccentricity (adopted)	    	    	    & 	$e$ &...& 0 \\
Orbital eccentricity (3-$\sigma$ upper-limit) &... &...& $0.060$\\
Projected stellar rotation velocity$^\dagger$ &   \vsini & km s$^{-1}$ & $9.89\pm0.48$\\
Sky-projected obliquity		    &   $\lambda$ & $^\circ$ \medskip & $20.1\pm9.7$\\
Stellar mass	    	    	    	    & 	$M_{\rm *}$ & $M_{\rm \odot}$ & $1.559\pm0.070$\\
Stellar radius	    	    	    	    & 	$R_{\rm *}$ & $R_{\rm \odot}$ & $2.26\pm0.17$\\
log (stellar surface gravity)     	    & 	$\log g_{*}$ & (cgs) & $3.920\pm0.050$\\
Stellar density     	    	    	    & 	$\rho_{\rm *}$ &$\rho_{\rm \odot}$ & $0.135\pm0.024$\\
Planet mass 	    	    	    	    & 	$M_{\rm P}$ &$M_{\rm Jup}$ & $2.242\pm0.080$\\
Planet radius	    	    	    	    & 	$R_{\rm P}$ &$R_{\rm Jup}$ & $1.46\pm0.13$\\
log (planet surface gravity)     	    & 	$\log g_{\rm P}$ & (cgs) & $3.378\pm0.059$\\
Planet density	    	    	    	    & 	$\rho_{\rm P}$ &$\rho_{\rm J}$ & $0.71\pm0.16$\\
Scaled orbital major semi-axis     &   $a/R_{\rm *}$ &...& {4.39}$\pm${0.27}\\
Orbital major semi-axis     	    	    & 	$a$ &AU  & $0.04619\pm0.00069$\\
Planet equilibrium temperature (uniform heat redistribution)	    	    & 	$T_{\rm P, A=0}$ &K & $2049\pm73$\\
\hline 
\end{tabular} \\ 
$\dagger$ We impose a Bayesian prior of  \vsini$ = 9.4 \pm 0.5$~km~s$^{-1}$ obtained from our spectral analysis (Sec. \ref{sec:spec}).\\
The following constant values are used: AU $= 1.49598\times10^{11}$~m; $R_{\rm \odot} = 6.9599\times10^8$~m; $M_{\rm \odot} = 1.9892\times10^{30}$~kg;\\
$R_{\rm Jup} = 7.1492\times10^7$~m; $M_{\rm Jup} = 1.89896\times10^{27}$~kg; $\rho_{\rm J} = 1240.67$~kg m$^{-3}$. \\ 
\end{table*} 

\begin{table}
\centering
\caption{Limb-darkening coefficients}
\label{tab:limb}
\begin{tabular}{cccccc} \hline 
Claret band & Light curves & $a_1$ &  $a_2$  &  $a_3$ & $a_4$  \\
\hline 
Cousins $R$  & WASP & 0.559 &  0.115  & 0.217  & -0.167 \\
Sloan $z^\prime$        & TRAPPIST  & 0.658 & -0.281  & 0.494  &-0.258 \\
\hline 
\end{tabular} 
\end{table} 

\section{Results and Discussion}

\subsection{Basic system parameters}

We find that WASP-71 is an evolved star, no longer on the main sequence, with mass and radius significantly larger than the Sun ($1.559\pm0.070$~\msol and $2.26\pm0.17$~\rsol respectively). The planet WASP-71b is 2.2 times more massive than Jupiter and has a radius 1.5 times that of Jupiter. WASP-71b joins a large number of giant planets which have radii larger than those theoretically predicted (by e.g. \citealt{BLL}). The radius of WASP-71b ($1.46\pm0.13$~\rjup) agrees well with the empirical radius calibration model of \cite{Enoch12}, which predicts a radius of $1.42\pm0.37$~\rjup (taking into account the uncertainties on our measured parameters, and the scatter in the empirical fit). The fit is based on the planet mass and equilibrium temperature and the stellar metallicity; in calculating this predicted radius, we assume a circular orbit for WASP-71b, and therefore omit the tidal-heating term from Equation~(10) of \cite{Enoch12}.

\subsection{Stellar spin - orbit alignment}

We make a highly-significant detection of the Rossiter-McLaughlin effect in the WASP-71 system; an analysis consisting of a purely Keplerian fit to the radial velocities produces a much larger $\chi^2$ fit statistic, and an F-test of the null hypothesis that the RM effect is not detected results in $P(F) \ll 0.0001$.

Our best-fitting value for the sky-projected spin-orbit obliquity is $\lambda = 20.1 \pm 9.7$~degrees. WASP-71 therefore follows the pattern observed by \cite{Winn10}, in that the system appears to be close to aligned ($|\lambda|$ is not measured to be greater than $10^\circ$ with at least a 3-$\sigma$ confidence), and the star is cool ($\teff < 6250$~K). The spin-orbit angle, $\lambda$, is only a projection of the true obliquity angle; only a measurement of the stellar rotation axis angle could reveal whether the planet's orbit is more inclined along the line-of-sight. Possible means of doing this include searching for mode splitting using asteroseismology \citep{Chaplin13}, or observing multiple spot-crossing events \citep{SOandWinn11}.

As \cite{triaud11} pointed out, stars with masses larger than 1.2~\msol will have started their existence on the zero-age main sequence at temperatures greater than 6250~K, the temperature above which \cite{Winn10} note a lack of aligned systems. Given a mass of $1.56\pm0.07$~\msol, WASP-71 will have been hotter than 6250~K in its recent past, and therefore presumably part of an isotropic distribution in obliquity.

If the system was originally significantly misaligned, and \cite{Winn10} are correct, then the planet may have become realigned since the star cooled and its convective zone grew. Alternatively, the orbit of WASP-71b may never have been significantly misaligned with the stellar rotation axis. We also note that the age of WASP-71 is around 2.5~Gyr, which is the limit older than which \cite{triaud11} observe that systems appear to be aligned.

\begin{acknowledgements}

WASP-South is hosted by the SAAO and SuperWASP by the Isaac Newton Group and the Instituto de Astrof\'{i}sica de Canarias;  we gratefully acknowledge their ongoing support and assistance. Funding for WASP comes from consortium universities and from the UK's Science and Technology Facilities Council (STFC).  TRAPPIST is a project funded by the Belgian Fund for Scientific Research (FNRS) with the participation of the Swiss National Science Foundation. MG and EJ are FNRS Research Associates. We thank the anonymous referee, and the editor, Tristan Guillot, for their thorough comments that led to improvements in the paper.
\end{acknowledgements}

\bibliographystyle{aa}
\bibliography{iau_journals,refs2}

\end{document}